\documentclass[prb,aps,amssymb,nofootinbib,floatfix,eqsecnum,showpacs]{revtex4}
\usepackage{graphicx,amsmath}

\begin{document}

\title{Strong coupling analysis of QED$_3$ for 
excitation spectrum broadening 
in undoped high-temperature superconductors}


\author{T. Morinari}
\email[]{morinari@yukawa.kyoto-u.ac.jp}
\affiliation{Yukawa Institute for Theoretical Physics, Kyoto
University, Kyoto 606-8502, Japan}


\date{\today}

\begin{abstract}
Theory of quantum electrodynamics in three spatial-time dimension is applied to
the two-dimensional $S=1/2$ quantum Heisenberg antiferromagnet
in order to investigate a doped hole in high-temperature superconductors.
Strong coupling analysis of the U(1) gauge field interaction
is carried out to describe spectral broadening observed in the undoped compounds.
It is found that the fermionic quasiparticle spectrum is of Gaussian form
with the width about $3J$, with $J$ being the superexchange interaction energy.
The energy shift of the spectrum is on the order of the quasiparticle band width,
which suggests that the system is in the strong coupling regime with respect to
the gauge field interaction describing the phase fluctuations about the 
staggered flux state.
\end{abstract}


\maketitle

\section{Introduction}
One of the most fundamental questions about high-temperature superconductivity
is how to describe doped holes introduced in the CuO$_2$ plane.
The simplest way to approach this problem is to investigate 
the single hole doped system.
Experimentally, the excitation spectrum associated with a single hole 
is observed by angle resolved photoemission spectroscopy (ARPES) 
in the undoped compounds such as Sr$_2$CuO$_2$Cl$_2$\cite{Wells1995,LaRosa1997}
and Ca$_2$CuO$_2$Cl$_2$.\cite{Ronning1998}
Although the spectrum is not sharp but quite broad, whose width is
ranging from $0.1$eV to $0.5$eV,
the experiments show that the trace of the peak indicates a dispersion whose maxima
are at $(\pm \pi/2, \pm \pi/2)$.
The dispersion near these points is quadratic and almost isotropic.
The band width is $2.2J \simeq 270{\rm meV}$ with $J$ the superexchange interaction.
This band width is much smaller than the band-structure estimation of $8t \simeq 2.8{\rm eV}$.
Furthermore, the observed spectra are not described by a conventional Lorentzian form 
but described by a Gaussian form.\cite{Shen2004}
These observations suggest that the quasiparticle excitations
in the undoped cuprates are quite different from conventional Fermi liquid quasiparticles.

In the slave-fermion theory of the $t$-$J$ model,
the single hole system has been analyzed 
by the self-consistent Born approximation.\cite{KaneLeeRead1989}
The effect of the spin-wave excitations is included in the self-energy 
in a self-consistent manner with omitting vertex corrections.
The resulting hole dispersion has minima at $(\pm \pi/2, \pm \pi/2)$,
and the band width is scaled by $J$.
The dispersion along $(0,0)$ to $(\pi,\pi)$ is in good agreement with 
the experiments.
Quantum Monte Carlo simulations based on a model, in which canonically transformed
spinless fermions propagate with antiferromagnetic spin correlation background,
like the slave-fermion formalism, are consistent 
with this result.\cite{BrunnerAssaadMuramatsu2000}
However, the dispersion along $(\pi,0)$ to $(0,\pi)$ is much smaller
than that in the experiments.
This discrepancy is improved by including the next nearest neighbor and
the third nearest neighbor hopping terms.\cite{TohyamaMaekawa2000}
This suggests that within this approach 
the quadratic behavior near the $(\pi/2,\pi/2)$ point along $(0,\pi)$ to $(\pi,0)$ 
has a different origin from that along $(0,0)$ to $(\pi,\pi)$.
Furthermore, it turns out that the damping effect due to the coupling 
with the spin-wave modes does not lead to a broad line shape.\cite{BalaOlesZaanen1995}
Recently Mischenko and Nagaosa studied a coupling to an optical phonon. 
\cite{MishchenkoNagaosa2004}
They numerically summed over Feynman diagrams including vertex corrections 
for phonons. 
It was argued that the coupling is in the strong coupling regime,
and so the quasiparticle spectrum becomes broad.
In this scenario the most enigmatic feature of the hole spectral broadening
in the undoped compounds is associated with phonon effects.
The dominant role is not played by the antiferromagnetic correlations 
which is believed to be essential for the mechanism of high-temperature superconductivity.

Here I take a different approach.
I consider the staggered flux state proposed 
in the literature \cite{AffleckMarston1988,WenLee1996}
from a mean field theory of the $S = 1/2$ antiferromagnetic Heisenberg model 
based on a fermionic representation of the spins.
The dispersion of the quasiparticle in the staggered flux phase is 
in good agreement with the experimentally obtained dispersion 
as pointed out by Laughlin.\cite{Laughlin1997}
Including phase fluctuations about the mean field,
the effective theory is described by quantum electrodynamics in 
three spatial-time dimension, which is called QED$_3$.
At mean field level, the fermions are massless.
By including the effect of the gauge field,
the mass of the Dirac fermions is induced.\cite{KimLee1999}
This mass is associated with the staggered magnetization.
The presence of the mass term is also suggested 
by a variational Monte Carlo approach.\cite{Hsu1990}
The quadratic dispersion around $(\pi/2,\pi/2)$ observed in the experiments
is consistent with the massive quasiparticle spectrum.
Furthermore, the quasiparticle dispersion is isotropic at $(\pi/2,\pi/2)$.

The purpose of this paper is to argue that 
the coupling of the fermions with the phase fluctuations 
leads to a broad Gaussian spectrum.
The QED$_3$ action with the mass term is analyzed
by performing a canonical transformation.
The spectral function is obtained by calculating
the Green's function of the Dirac fermion which is associated with 
a single quasiparticle propagation.
It is shown that the spectral function shows a broad Gaussian peak 
whose width is on the order of $J$.

The rest of the paper is organized as follows. 
In Sec.\ref{sec_formalism}, we describe the QED$_3$ formalism of the staggered flux state.
After taking the transverse gauge, a canonical transformation is applied.
The quasiparticle Green's function is computed in Sec.\ref{sec_results}.
The spectral function is obtained with including vertex functions
arising from random phase approximation about the instantaneous 
longitudinal gauge field interaction.
Implications of the result is discussed in \ref{sec_discussion}.

\section{QED$_3$ theory of the staggered flux state\label{sec_formalism}}
For the description of the $S=1/2$ two-dimensional quantum Heisenberg antiferromagnet,
I take the following QED$_3$ action in the real time formalism as the effective theory:
\begin{equation}
S = \int {d^3 x} \left[ {\overline \psi  \left( x \right)\left[ 
{i\gamma ^\mu  \left( {\partial _\mu   - ia_\mu  } \right) - m\sigma _3 } \right]\psi 
\left( x \right) - \frac{1}{{4e_a ^2 }}f_{\mu \nu } f^{\mu \nu } } \right],
\label{eq_QED3}
\end{equation}
where the Dirac fermion fields, $\psi (x)$, consist of four component associated with
even and odd sites and two independent nodes.
Due to the spin degrees of freedom there are two species of $\psi (x)$.
(For the derivation, see Appendix \ref{app_QED3}.)
Hereafter the spin index for $\psi (x)$ is suppressed because
the spin degrees of freedom does not play an important role in the following 
calculation.
The action (\ref{eq_QED3}) describes low-lying excitations around $(\pm \pi/2, \pm \pi/2)$
because the above continuum model was derived by taking the continuum limit 
at those points.
Note that the theory is particle-hole symmetric.
Therefore, the quasiparticle properties are identical to the quasihole properties.

For the gauge, I take the transverse gauge: $\nabla \cdot {\bf a} = 0$.
In this gauge, the interaction between the fermions mediated by the longitudinal
part of the gauge field is instantaneous as in the conventional electromagnetic field formulation:
\begin{eqnarray}
S &=& \int {d^3 x} \overline \psi  \left( x \right)
\left[ {i\gamma ^0 \partial _t  + i\gamma ^j 
\left( {\partial _j  + ia_j } \right) - m\sigma _3 } \right]\psi \left( x \right)
\nonumber \\
& & + \frac{{e_a^2 }}{{4\pi }}\int {d^3 x} \int {d^2 {\bf{r'}}} \left[ {\rho \left( {{\bf{r}},t} \right) 
- \rho _0 } \right]\left[ {\rho \left( {{\bf{r'}},t} \right) - \rho _0 } \right]\ln \left| {{\bf{r}} 
- {\bf{r'}}} \right| \nonumber \\
& &  + \frac{1}{{2e_a ^2 }}\int {d^3 x} \left[ {\left( {\partial _t {\bf{a}}} \right)^2  
- \left( {\nabla  \times {\bf{a}}} \right)^2 } 
\right],
\end{eqnarray}
where the background gauge charge, $-e_a \rho_0$, comes from the constraint on the fermion number
to represent the spin $1/2$.
In three spatial-time dimension, the "Coulomb" interaction is described by
$V\left( r \right) =  - \frac{{e_a^2 }}{{2\pi }}\ln r$.
Under the transverse gauge, the vector potential is represented by
\begin{equation}
 a_x \left( q \right) =  - \frac{{iq_y }}{q}a\left( q \right),
\hspace{2em}
 a_y \left( q \right) = \frac{{iq_x }}{q}a\left( q \right),
\end{equation}
in the momentum space.
Quantizing the transverse gauge field, the Hamiltonian is
\begin{eqnarray}
H &=& \int {d^2 {\bf{r}}} \overline \psi  \left( {\bf{r}} \right)\left( { - i\gamma ^j \partial _j  
+ m\sigma _3 } \right)\psi \left( {\bf{r}} \right) 
+ \int {d^2 {\bf{r}}} \sum\limits_q 
    e^{iq \cdot {\bf{r}}} \overline \psi  
    \left( {\bf{r}} \right)
    \frac{i}{q}
    \left( {q_x \gamma _y  - \gamma _x q_y } 
    \right)
\sqrt{\frac{{e_a^2 }}{{2\omega _q }}} \left( {b_q  + b_{ - q}^\dag  } \right)
\psi \left( {\bf{r}} \right)
\nonumber \\
& & + \frac{1}{2}\int {d^2 {\bf{r}}} \int {d^2 {\bf{r'}}} 
V(\left| \bf{r} - \bf{r}' \right| )
\left[ {\rho \left( {\bf{r}} \right) - \rho _0 } \right]
\left[ {\rho \left( {{\bf{r'}}} \right) - \rho _0 } \right]
\nonumber \\
& & + \sum\limits_q {\frac{{\omega _q }}{2}\left( {b_q b_q^\dag   + b_q^\dag  b_q } \right)}.
\label{eq_H}
\end{eqnarray}

In order to investigate the strong coupling effects, I perform the following canonical 
transformation\cite{Mahan}:
\begin{equation}
 \overline H  = e^s He^{ - s},
\end{equation}
where
\begin{equation}
 s = \int {d^2 {\bf{r}}} \sum\limits_q {\left( {b_q  - b_{ - q}^\dag  } \right)\psi ^\dag  \left( {\bf{r}} \right)
M_q \left( {\bf{r}} \right)\psi \left( {\bf{r}} \right)}.
\end{equation}
(The area of the system is set to unity.)
This is a unitary transformation if $
M_{ - q} ^\dag  \left( {\bf{r}} \right) = M_q \left( {\bf{r}} \right)$.
The function $M_q({\bf r})$ is chosen so that the interaction term 
between the Dirac fermions and the gauge field is cancelled by
$[s,H]$:
\begin{equation}
M_q \left( {\bf{r}} \right) =  - \sqrt {\frac{{e_a^2 }}{{2\omega _q ^3 }}} 
e^{iq \cdot {\bf{r}}} \frac{i}{q}\gamma _0 \left( {q_x \gamma _y  - \gamma _x q_y } \right).
\end{equation}
Under this canonical transformation, the fermion fields transform as
\begin{equation}
e^s \psi \left( {\bf{r}} \right)e^{ - s}  = e^{X\left( {\bf{r}} \right)} \psi \left( {\bf{r}} \right).
\end{equation}
The function $X\left( {\bf{r}} \right)$ is,
\begin{equation}
X\left( {\bf{r}} \right) =  - \sum\limits_q {\left( {b_q  - b_{ - q}^\dag  } \right)
M_q \left( {\bf{r}} \right)}. 
\end{equation}

\section{\label{sec_results}Quasiparticle Green's function}
Now I compute the time-ordered Green's function:
\begin{eqnarray}
G\left( {{\bf{r}}, t} \right) &=&  
- i \left\langle {T  \psi \left( {{\bf{r}}, t } \right)\psi ^\dag  
\left( {0,0} \right)} \right\rangle
\nonumber \\
&=&  
- i \left\langle {T  
{\rm e}^{X({\bf r},t)}
\psi \left( {{\bf{r}}, t } \right)\psi ^\dag  
\left( {0,0} \right)} 
{\rm e}^{-X({\bf 0},0)}
\right\rangle_{\overline{H}}.
\label{eq_G}
\end{eqnarray}
This Green's function has a matrix form of $4\times 4$.
But the matrix is divided into two blocks
in which each part describes either 
the Dirac fermion fields near $(\pi/2,\pi/2)$ or
the Dirac fermion fields near $(-\pi/2,\pi/2)$.
It is enough to focus on one of them because 
two components are decoupled as far as long-wave length
gauge field fluctuations are concerned.
The superscript $(1)$ is used to denote the former.
Diagonalizing the factor with $\gamma$ matrices, $X({\bf r})$ is
\begin{equation}
X^{(1)}\left( {\bf{r}} \right) = i\sum\limits_q 
{\left( {b_q e^{iq \cdot {\bf{r}}}  + b_q^\dag  e^{ - iq \cdot {\bf{r}}} } \right)
\sqrt {\frac{{e_a^2 }}{{2\omega _q ^3 }}} 
U_q \tau_3 {U_q ^\dag  }
},
\end{equation}
where 
\begin{equation}
U_q = \frac{1}{\sqrt{2}} \left(
\begin{array}{cc}
1 & -(q_x - i q_y)/q \\
(q_x + iq_y)/q & 1 
\end{array}
\right).
\end{equation}
The first term in Eq.(\ref{eq_H}),
which does not change its form by
the canonical transformation, 
is diagonalized by a unitary transformation as well.
Finally, by making use of the following formula,
\begin{equation}
\left\langle {e^{Ab^\dag   + Bb} e^{Cb^\dag   + Db} } 
\right\rangle  = e^{\frac{1}{2}\left( {AB + CD + 2BC} \right)} 
e^{\left( {A + C} \right)\left( {B + D} \right)n\left( \omega  \right)},
\end{equation}
for bosons, Eq.(\ref{eq_G}) is
\begin{equation}
G^{\left( 1 \right)} \left( {{\bf{r}},t } \right) 
=  - i \sum\limits_k {e^{i{\bf{k}} \cdot {\bf{r}}} } 
e^{ - iE_k t} R_k
\exp \left[ 
{ 
- K({\bf r},t)
} \right],
\end{equation}
at $T=0$.
The retarded Green's function has the same form for $t>0$.
Here,
\begin{equation}
R_k = \prod_{\bf q} 
\left[ 
   \frac{1}{4}
    \left( 
        1+\frac{k_x q_y - k_y q_x}{q \sqrt{k^2+m^2}} 
    \right)
\right],
\end{equation}
\begin{equation}
K({\bf r},t) = \sum\limits_q {\frac{{e_a^2 }}{{2\omega _q ^3 }}
  \left( {1 - e^{ - i\omega _q t} e^{i{\bf{q}} \cdot {\bf{r}}} } \right)}.
\end{equation}
Performing the Fourier transform, 
the spectral function $A^{(1)}({\bf k},\omega)$ is given by
$A^{(1)}({\bf k},\omega) = -\frac{1}{\pi} {\rm Im} 
G^{(1)} ({\bf k}, \omega)$.

So far the bare vertex function has been used for the computation.
However, in the long-wave length limit taking the bare vertex is not appropriate
as manifestly seen by infrared divergence in $K({\bf r},t)$.
For the vertex part, random phase approximation is applied with respect to 
the longitudinal interaction term.
The bare vertex is reduced by the factor of $1/(1-\pi_q)$, where
\begin{equation}
\pi_q = v_q \sum\limits_k {Tr\left[ {G_k \gamma _0 G_{k + q} \gamma _0 } \right]}
=  - \frac{{me_a^2 }}{\pi }\frac{1}{{q^2 }} + O(1),
\end{equation}
with $v_q = e_a^2/q^2$.
For this computation, it is convenient to use the Euclidean formalism 
because the main contribution comes from $\pi_q$ with $q = ({\bf q},0)$,
where the Minkowskiian formalism leads to the same result.

Including the vertex correction the function $K({\bf r},t)$ is,
\begin{equation}
K({\bf r},t) = \frac{e_a^2}{2} \int_0^{\Lambda} dq \frac{1}{q^2}
\left( \frac{1}{1-\pi_q} \right)^2
\left[ 1-{\rm e}^{-i\omega_q t} J_0 (qr) \right],
\label{eq_K}
\end{equation}
with $J_0(x)$ the zero-th order of the Bessel function of the first kind.
The ultraviolet cutoff $\Lambda$ introduced here because
the wave length of fluctuations is larger than the lattice constant.
The integrand is expanded with respect to $q$, before the integration.
The result is
\begin{equation}
K({\bf r},t) \simeq iE_s t + \frac{1}{8} \Delta^2 r^2  
+ \frac{1}{4} \Delta^2 t^2,
\end{equation}
where
\begin{equation}
E_s = 
\frac{{e_a^2 }}{4}\log \left( {\frac{{\pi \Lambda ^2 }}{{eme_a^2 }}} \right),
\end{equation}
\begin{equation}
\Delta^2 = \frac{{e_a^2 \Lambda }}{2\pi}.
\end{equation}
The first term in Eq.(\ref{eq_K}) represents the energy shift, $E_s$.
The shape is changed to a broad Gaussian form by the subsequent terms
as shown below.
For the factor $R_k$, the analytic expression was not obtained.
From a numerical computation I found that 
$R_k$ is linear in $k$ at $0<k<k_c$, with $k_c \simeq 1$,
and reaches a saturated value of $0.2$ for $k > k_0$.
To approximate $R_k$, I took an approximate form of $R_k \simeq 0.2k$.
In computing the Fourier transform, it is useful to note that
the integration with respect to ${\bf r}$ 
and that with respect to $t$ are carried out separately.
For $k\ll \Delta$, I obtained
\begin{equation}
A^{(1)}({\bf k},\omega)
\simeq \frac{0.10}{\sqrt{8\pi^2}}
\exp \left[ 
-\frac{(\omega-E_k-E_s)^2}{\Delta^2}
\right].
\end{equation}
Therefore, the Dirac fermion energy spectrum is shifted by $E_s$ and
is a Gaussian form with the width of $\Delta$.
For the estimation of these values, the spin wave velocity is assumed to be, 
$c_{sw} = 1.6J$.
The mass term is evaluated as $m\simeq 1.3J$ from the ARPES experiments 
by fitting the dispersion near $(\pi/2,\pi/2)$.
The gauge charge is simply taken from the factor of the Maxwellian term
obtained by integrating out the Dirac fermions: $e_a^2 = 3\pi m$.
(If the same calculation is carried out for Dirac fermions with $k>k_0$,
then the gauge charge value is $e_a^2 \simeq 3\pi^2 k_0/4$, for $k_0 \gg m$.
Therefore, the above choice is the minimum value for the gauge coupling.)
Recalling the fact that $c_{sw}$ is taken unity in the above calculation,
I found $E_s \simeq 2J$ and $\Delta \simeq 3J$.
This value of $\Delta$ is consistent with the above assumption about $k$.
The weight of the Gaussian spectrum is $\sim 0.05$ for this value of $\Delta$.
Because of the spin degrees of freedom and the degenerate nodes, 
the weight is $\sim 0.2$ in total.
From the value of $E_s$, 
one can get insight about the strength of the coupling.
If the system is in the (weak) strong coupling, 
the value of $E_s$ is expected to be large (small).
The fact that $E_s$ is on the order of the band width
suggests that the coupling is in the strong coupling regime.

\section{\label{sec_discussion}Discussion}
In this paper, the spectral function of the quasiparticle 
in the staggered flux state with phase fluctuations
was computed within the effective theory described by the QED$_3$.
It was shown that the quasiparticle spectra become a broad Gaussian form
with an energy shift due to the gauge field interaction.
The estimated spectrum width is consistent with the experiments.
The analysis suggests that the coupling to the gauge field 
is in the strong coupling regime.

Since the model is based on the continuum approximation,
the result is applicable to the quasiparticle excitations near 
$(\pm \pi/2,\pm \pi/2)$.
However, the result is extended to other ${\bf k}$ points
by formulating the theory on the lattice.
Such a model is useful to study the change of the width of the spectra
away from $(\pm \pi/2,\pm \pi/2)$.

As for the vertex correction, random phase approximation
is applied with respect to the interaction arising from the longitudinal
component of the gauge field.
Of course, this is not the full vertex correction.
In the long-wave length limit, there are other intermediate
processes.
However, it is expected that dominant contribution is covered by the above 
vertex correction because the longitudinal component plays 
a major role in screening the gauge charge.

Finally, let me comment on vanishing quasiparticle spectra
observed in the experiments \cite{Wells1995,LaRosa1997}
along the line from $(\pi/2,\pi/2)$ to $(\pi,\pi)$
and that from $(\pi,0)$ to $(\pi,\pi)$. 
A similar behavior is also observed in the pseudogap phase,\cite{ShenRMP}
that is, only a part of the Fermi surface is observed as an arc shape.
\cite{Norman1998}
One might expect that damping effect coming from
the coupling to the gauge field leads to the suppression
of the quasiparticle peaks.
However, it turns out that the inclusion of a slight hopping
term in the staggered flux state leads to the vanishing of the spectrum
in the second magnetic Brillouin zone.
I will discuss this matter in a future publication.

\acknowledgments
I would like to thank Prof. T. Tohyama for helpful discussion.
This work was supported by Grant-in-Aid for Young Scientists
and the 21st Century COE "Center for Diversity and
Universality in Physics" from the Ministry of Education, Culture,
Sports, Science and Technology (MEXT) of Japan. 
The numerical calculations were carried out in part 
on Altix3700 BX2 at YITP in Kyoto University.

\appendix
\section{Derivation of QED$_3$ action\label{app_QED3}}
In this appendix, I derive the QED$_3$ action as the effective theory
for the $S=1/2$ two-dimensional Heisenberg antiferromagnet:
\begin{equation}
H=J\sum_{\langle i,j \rangle} {\bf S}_i \cdot {\bf S}_j.
\end{equation}
A fermion representation is introduced for the spin $1/2$:
${\bf S}_{j\mu} = f_{j\alpha}^{\dagger} \sigma_{\alpha \beta} f_{j\beta}/2$
($\mu=x,y,z$).
$\sigma^{\mu}$ are the Pauli spin matrices.
These fermions need to satisfy the constraint, 
$\sum_{\alpha} f_{j\alpha}^{\dagger} f_{j\alpha} = 1$.
Introducing Lagrange multipliers to take into accout the constraint, 
the Hamiltonian is,
\begin{equation}
H =  - \frac{1}{2}J\sum\limits_{\left\langle {i,j} \right\rangle } 
{f_{i\alpha }^\dag  f_{j\alpha } f_{j\beta }^\dag  f_{i\beta } }  
+ \sum\limits_j {\lambda _j \left( {f_{j\sigma }^\dag  f_{j\sigma }  - 2S} \right)},
\end{equation}
up to a constant term.
The mean field taken in the $\pi$-flux state is 
$\chi _{ij}  = \left\langle {f_{j\alpha }^\dag  f_{i\alpha } } \right \rangle$.h,
by choosing a suitable gauge.\cite{LeeNagaosaWen2006}
Since the system is homogeneous, uniform $\chi_{ij}$ and $\lambda_j$ are assumed:
$\chi_1 = \chi_{j+\hat{x},j}$,
$\chi_2 = \chi_{j,j-\hat{y}}$,
$\chi_3 = \chi_{j-\hat{x},j}$,
and
$\chi_4 = \chi_{j,j+\hat{y}}$,
with $j$ residing at an even site.
Numerically solving the mean field equations for $\chi_j$ $(j=1,2,3,4)$, with 
setting $\lambda=0$,
the $\pi$-flux state\cite{AffleckMarston1988} is found
in which $\chi_1 \chi_2 \chi_3 \chi_4/|\chi_1 \chi_2 \chi_3 \chi_4| = -1$.

The quasiparticle energy dispersion in the staggered flux state is 
\begin{equation}
E_k = \pm \frac{J}{2} | 
\chi_1 {\rm e}^{-ik_x} + 
\chi_2^* {\rm e}^{ik_y} + 
\chi_3 {\rm e}^{ik_x} + 
\chi_4^* {\rm e}^{-ik_x}|.
\end{equation}
$|E_k|$ has minima at $(\pm \pi/2,\pm \pi/2)$,
and around these points the energy dispersion 
has the relativistic form.
Introducing the even and odd site fields, 
$f_{ek\alpha} = (f_{k\alpha} + f_{k+Q,\alpha})/\sqrt{2}$
and
$f_{ok\alpha} = (f_{k\alpha} - f_{k+Q,\alpha})/\sqrt{2}$
with $Q=(\pi,\pi)$,
and expanding $E_k$ around $(\pm \pi/2, \pi/2)$,
the Hamiltonian is rewritten as,
\begin{eqnarray}
 H &\simeq & J{\sum_k}^{\prime} 
   {\left( {\begin{array}{*{20}c}
   {f_{e1k\alpha }^\dag  } & {f_{o1k\alpha }^\dag  }  \\
\end{array}} \right)\left( {\begin{array}{*{20}c}
   0 & { - \chi _1 ^* k_x  + \chi _2 k_y }  \\
   { - \chi _1 k_x  + \chi _2 ^* k_y } & 0  \\
\end{array}} \right)} \left( {\begin{array}{*{20}c}
   {f_{e1k\alpha } }  \\
   {f_{o1k\alpha } }  \\
\end{array}} \right)
\nonumber \\
& &   + J{\sum_k}^{\prime} {\left( {\begin{array}{*{20}c}
   {f_{e2k\alpha }^\dag  } & {f_{o2k\alpha }^\dag  }  \\
\end{array}} \right)\left( {\begin{array}{*{20}c}
   0 & {\chi _1 ^* k_x  + \chi _2 k_y }  \\
   {\chi _1 k_x  + \chi _2 ^* k_y } & 0  \\
\end{array}} \right)} \left( {\begin{array}{*{20}c}
   {f_{e2k\alpha } }  \\
   {f_{o2k\alpha } }  \\
\end{array}} \right).
\end{eqnarray}
Here the indices $1$ and $2$ are introduced to
denote the fields around $(\pi/2,\pi/2)$ and 
those around $(-\pi/2,\pi/2)$.
The summation with respect to $k$ is taken over the magnetic 
Brillouin zone: $|k_x \pm k_y|<\pi$.
Choosing $\chi_1 = \chi_3 = i|\chi|$
and $\chi_2 = \chi_4 = |\chi|$,
and setting $\psi_{k\alpha}^{\dagger} = 
\left( f_{e1k\alpha}^{\dagger},f_{o1k\alpha}^{\dagger},
f_{o2k\alpha}^{\dagger},f_{e2k\alpha}^{\dagger} \right)$,
the action is, in the continuum limit,
\begin{equation}
S = \int {d^3 x} \overline \psi  \left( x \right)
i\gamma ^\mu  \partial _\mu  \psi \left( x \right),
\end{equation}
where $\overline{\psi} = \psi^{\dagger} \gamma^0$ and
the $\gamma$ matrices are
\[
\gamma ^0  = \left( {\begin{array}{*{20}c}
   {\tau _3 } & 0  \\
   0 & { - \tau _3 }  \\
\end{array}} \right), 
~~
\gamma ^1  = \left( {\begin{array}{*{20}c}
   {i\tau _1 } & 0  \\
   0 & { - i\tau _1 }  \\
\end{array}} \right), 
~~
\gamma ^2  = \left( {\begin{array}{*{20}c}
   {i\tau _2 } & 0  \\
   0 & { - i\tau _2 }  \\
\end{array}} \right).
\]

Phase fluctuations about the staggered flux mean field state
are included by the U(1) gauge field, $a_{\mu}$,
by replacing $\partial_{\mu}$ with $\partial_{\mu} - ia_{\mu}$.
Integrating out the high-energy Dirac fermion fields with 
$k > k_0$ leads to the dynamics of the gauge field,
which has the Maxwellian form.
Numerically solving the Schwinger-Dyson equation,
it is found that the self-energy has a non-zero mass.\cite{KimLee1999}
Physically this mass is associated with the presence of 
the staggered magnetization which is absent 
at the mean field level.
From the variational Monte Carlo approach with
the finite mass $m$, it is shown that the mean field 
energy improves by the inclusion of $m$.\cite{Gros1989,LeeFeng1988}
The action (\ref{eq_QED3}) is obtained by including the mass term
arising from the staggered magnetization.


\bibliography{../refs/tmrefs}

\end{document}